\documentclass[aps,prb,twocolumn,amsmath,amssymb,superscriptaddress]{revtex4}

\usepackage{amssymb}
\usepackage{amsmath}
\usepackage{graphicx}
\usepackage{textcomp}
\usepackage{color}
\usepackage{amsfonts}
\usepackage{epsfig}
\usepackage{hyperref}
\usepackage{bm}
\usepackage[caption=false]{subfig} 
\usepackage{booktabs}
\usepackage{tabularx}
\usepackage{array}
\usepackage{bbold}

\newcommand{\arctanh}[1]{\operatorname{arctan}}

\newcolumntype{M}[1]{>{\centering\arraybackslash}m{#1}}

\DeclareMathAlphabet\mathbfcal{OMS}{cmsy}{b}{n}

\begin{document}

\author{Reena Gupta}
\author{Andrea Droghetti}
\email{andrea.droghetti@tcd.ie}
\affiliation{School of Physics and CRANN, Trinity College Dublin, Dublin 2, Ireland}
\date{\today}

\title{Current-induced spin polarization in chiral tellurium: a first-principles quantum transport study}

\begin{abstract}
Te is a naturally p-doped semiconductor with a chiral structure, where an electrical current causes the conduction electrons to become spin polarized parallel to the transport direction. In this paper, we present a comprehensive theoretical study of this effect, named CISP, by employing density functional theory (DFT) combined with the non-equilibrium Green's functions (NEGF) technique for quantum transport. CISP can be quantitatively described 
in terms of the non-equilibrium spin density, which we show to be localized around the atoms. We then compute the atomic magnetic moments as a function of doping and electronic temperature, obtaining results overall consistent with those of previous theoretical studies\cite{ts.pu.17,Roy2022}. Beyond that, our DFT+NEGF calculations show that CISP also leads to a spin current, which is found to be a quite large fraction of the charge current, indicating that Te can be an efficient material for spin transport. We further predict that the resistance along a ballistic Te wire changes when an external magnetic field is applied parallel or antiparallel to the direction of the charge current. The computed magnetoresistance ratio is however quite small ($\sim 0.025\%$). Finally, we conclude by arguing that CISP, as treated within the DFT+NEGF framework, coincides with the phenomenon called chiral-induced spin selectivity, recently reported in several nano-junctions. 
\end{abstract}

\maketitle
\section{Introduction}
Charge-spin conversion phenomena\cite{ot.sh.17,ha.ot.18,so.re.16}, which are mediated by the spin-orbit coupling (SOC), are central for present-day spintronics as they enable all-electrical generation, manipulation, and detection of spins in solid state devices. A prominent example is the current-induced spin polarization (CISP)\cite{ga.tr.19} and its reciprocal effect\cite{ga.iv.02}. 
A dc electrical current flowing through a non-magnetic material can lead to the spin-polarization of the conduction electrons and, conversely, a non-equilibrium spin polarization can induce an electrical current. At the microscopic level, the effect emerges because the out-of-equilibrium electrons' distribution of a driven system presents a momentum asymmetry, which, in presence of spin-momentum locking, leads an unbalance spin population\cite{ed.90}. \\

To date, CISP has been mostly studied in low dimensional systems, where it is commonly called the Rashba-Edelstein or inverse spin galvanic effect\cite{si.va.15}, and it has been theoretically described by means of effective 2D electron gas models\cite{ar.ly.89,ed.90, ma.zh.08,go.ma.17}. Experimentally, CISP has been reported in semiconducting \cite{ka.my.04,si.yu.04,si.my.05,ga.da.06, ya.he.06,ch.ch.07,no.tr.14} and van der Waals heterostructures \cite{gh.ka.19, li.zh.20}, on the surfaces of topological insulators\cite{an.ha.14,Kondou2016}, and the interfaces of heavy metal thin films\cite{mi.mo.11,Zhang2014,zh.ya.15}. Yet, the effect does not occur only in low dimensional systems, but can also emerge in bulk materials whenever the crystalline structure is chiral, or, more generally, gyrotropic (i.e., the symmetry is such that the components of polar and axial vectors are transformed in accordance with equivalent representations). In fact, CISP was first predicted over forty years ago for elemental tellurium \cite{1978JETPL}, a naturally hole-doped semiconductor which possesses strong SOC and a chiral structure. Signatures of the effect in Te were then found in early studies about the material's optical activity \cite{1979JETPL}, while the spin polarization was definitely probed via NMR measurements a few years ago \cite{Furukawa2017, PhysRevResearch.3.023111}. Recently, the interest for CISP in Te has been re-fueled by some transport experiments\cite{ri.av.19,Calavalle2022}. In particular, Calavalle {\it et al.} detected
a resistance which was dependent on the relative orientation of the current and of an external magnetic field in single-crystal nanowires\cite{Calavalle2022}. The effect, called unilateral magnetoresistance or  electrical magnetochiral anisotropy\cite{PhysRevLett.87.236602}, could be tuned up to a factor six by electrostatic gating. The results have opened the way towards the exploitation of CISP for magnet-free spintronic devices.\\

CISP has sometimes been discussed alongside with another intriguing effect, named chiral-induced spin selectivity (CISS) \cite{ra.an.99,xi.ma.11,na.wa.12, ge.ka.13, Dor2013,ar.me.17,Naaman2019,na.pa.20,li.wa.20}, which is observed in molecular junctions with chiral structures. In CISS, a collinear spin current is generated by driving a charge current through a non-magnetic molecule and is detected by means of a ferromagnetic electrode. Although the underlying physics remains debated and poorly understood \cite{ev.ah.22}, the number of reports about the effect has rapidly grown over the last few years, and the results qualitatively resemble those from the magnetotransport experiment by Calavalle {\it et al.} for Te.\\

On the theoretical side, accurate predictions of CISP can be obtained by means of first-principles electronic structure calculations. Different approaches based on Kohn-Sham Density Functional Theory (DFT) have been proposed and applied to both bulk materials\cite{te.ro.23} and surfaces\cite{PhysRevB.91.035403,dr.to.23}. In particular, Te was studied by Tsirkin {\it et al.} and Roy {\it et al.} in Refs. [\onlinecite{ts.pu.17}] and  [\onlinecite{Roy2022}], respectively. They employed DFT combined with the Kubo and the Boltzmann transport formalisms to estimate the CISP magnitude in terms of non-equilibrium atomic magnetic moments. Furthermore, a recent paper also derived a theory for the unilateral magnetoresistance in the ohmic transport regime\cite{liu2023electrical} relevant for the experiments by Calavalle {\it et al.} Yet, despite these advances, there are still many open questions about CISP, its actual magnitude, and how it depends on different material's parameters. Furthermore, the connection of CISP with CISS remains somewhat ambiguous as theoretical studies of the two effects typically rely on very different frameworks and assumptions.\\ %In this paper we address some of these issues. In particular the goal is to achieve quantitative description of CISP and to predcit whether a magnetochiral anisoptropy effectthe ballistic transport regime, potentially relevant for future experiments with nanoscale samples,.}\\

In this paper, we employ DFT combined with the non-equilibrium Green's function (NEGF) technique \cite{Ba.Mo.02,Ta.Gu.01,rocha2006spin} to study the emergence of CISP in Te at an accurate quantitative level. Our approach, called DFT+NEGF, is commonly used in the study of coherent transport in molecular junctions and point contact\cite{RevModPhys.92.035001}, but, to our knowledge, has never been applied to CISP. Here, we demonstrate its potential for achieving a description of the effect complementary, and somewhat more complete, with respect to that presented in previous works \cite{liu2023electrical,Roy2022}. Furthermore, we aim at predicting whether CISP can affect the magnetransport properties down to the ballistic transport regime, which can eventually be reached in future experiments. The calculations in this regard are also relevant in order to address the issue of comparing CISP and CISS within a common theoretical framework.\\ 

Our calculations give an accurate estimate of the CISP magnitude in terms of the non-equilibrium spin density, which we find to be localized around the atoms and warping around the Te chiral structure. We then compute the atomic magnetic moments as a function of doping and temperature, obtaining results consistent with those of the previous DFT studies in Refs. [\onlinecite{Roy2022, ts.pu.17}]. Moreover, we find that the CISP ``figure of merit'', i.e., the ratio of the spin density over the charge current density, is large and comparable to the one of heavy metal thin films, indicating that Te is an efficient material for charge-to-spin conversion.\\

Going beyond the magnetic properties, our DFT+NEGF calculations also show that there are spin currents emerging together with CISP. We can therefore establish a clear connection between the phenomenon and the material's magnetotransport properties. We predict that a unilateral magnetoresistance can appear not just in the ohmic transport regime, considered in previous works, but also in the ballistic one analyzed here. However, the computed magnetoresistance ratio is very small in this latter case. \\

Finally, our DFT+NEGF calculations for Te can be directly compared to those for the CISS effect in molecular junction \cite{zo.se.20,or.re.23, ga.bl.23,na.mu.23}. We find that all differences can be traced back to materials' properties and symmetry rather than to fundamental aspects. Hence, we conclude that CISP and CISS are the same effect when analyzed from the perspective of DFT+NEGF calculations.\\

The paper is organized as follows. We start in Sec. \ref{sec.TeandCISP} by describing the Te crystal structure and explaining how its symmetry allows for CISP. We then continue in Sec. \ref{sec.computational} by briefly reviewing the theoretical framework, and we provide the computational details of the DFT and DFT+NEGF calculations. The results are presented in Sec. \ref{sec.results}, where we describe the band structure (Sec. \ref{sec.bands}), the non-equilibrium spin density (Sec. \ref{sec.spin_density}), the CISP figure of merit (Sec. \ref{sec.figure_merit}), the spin current (Sec \ref{sec.spin_current}),  and the magnetoristance calculations (Sec. \ref{sec.MR}). Finally, we compare our work with recent computational studies for chiral molecular junctions in Sec. \ref{sec.discussion}, and we conclude in Sec. \ref{sec.conclusion}. 

\section{Te structure and CISP}\label{sec.TeandCISP}
At ambient conditions, Te has a trigonal crystal structure (Te-I) consisting of helical chains, which can be either right- or left-handed defining two enatiomers with a specific chirality. %providing a defined chirality to the system. 
%Its point is $D_3$ and %The three-fold screw symmetry $C_3$, responsible for the chirality, lies along the c-axis while a two-fold $C_2$ axis lie in the a-b plane. Van der Waals forces keep together adjacent atomic chains. The lattice constants employed are $a=4.51$ \AA ~and $c=5.96$ \AA.
The atoms within each chain are covalently bonded together, whereas adjacent chains are held together by van der Waals forces. The right-handed and left-handed structures are respectively described by the enantiomorphic space groups $P3_121$ and $P3_221$, sharing the point
group $\mathbb{D}_3$. The three-fold screw symmetry $\mathbb{C}_3$, responsible for the chirality, lies along the $c$ axis, while a two-fold $\mathbb{C}_2$ axis lie in the perpendicular direction. The primitive unit cell is hexagonal [see Fig. S1-a in the Supplementary Material (SM)], and we consider here the experimental lattice parameters $a=4.51$ \AA~and $c=5.96$ \AA, respectively in-plane and along the $c$ axis. Although Te is a narrow band-gap semiconductor\cite{Doi1970TheVB, Shalygin2012, Roy2022}, experimental samples are naturally p-doped and conductive because of Te vacancies\cite{na.ga.20}, and therefore they  show CISP. \\

At the phenomenological level, CISP is described by the equation \cite{ga.tr.19} $S^{a}=\gamma^{a}_{ b} j_{b}$ which linearly couples the non-equilibrium spin density $\mathbf{S}=(S^x,S^y,S^z)$ to the charge current density $\mathbf{j}=(j_x,j_y,j_z)$ via the components $\gamma^{a}_{b}$ of a second-rank pseudo-tensor $\pmb{\gamma}$, which is analogous to the gyration tensor\cite{Landau} determining the natural optical activity (note Einstein's convention of repeated indices). In case of Te, the $\mathbb{D}_3$ point symmetry dictates that $\pmb{\gamma}$ must be diagonal\cite{ts.pu.17}. Thus, the induced spin density is expected to be parallel to the current density direction. If the current flows along the $c$ axis, assumed to coincide with $z$ Cartesian direction, we will get $S^z=\gamma_z^z j_z$ with the sign of $\gamma^z_z$ being positive (negative) for left-(right-) handed Te. This behaviour is different from that observed in other well-studied systems, such as surfaces and heterostructures, where the flow of a charge current generally results in a perpendicularly oriented homogeneous spin density \cite{Zhang2014, dr.to.23}. For this reason, CISP in Te has been labeled as ``unconventional'' \cite{Calavalle2022}. In the following we will provide a microscopic and fully quantitative description of the phenomenon from first-principles. \\

%\begin{figure*}[!ht]
%    \centering
%    \includegraphics[width=0.9\textwidth]{Fig1.pdf}
%    \caption{\textcolor{blue}{figure to be changed with the real space density}(a) Typical systems studied in the DFT+NEGF calculations. It is subdivided into a central region (C), and left (L) and right (R) leads. The arrows rappresent the current induced magnetic moments. (b) Spin density profile of the central region obtained for a charge current density $j_z=2.7 \times 10^6$ A/cm$^2$}\label{fig.CISP}
%\end{figure*}

\begin{figure*}[!ht]
    \centering
    \includegraphics[width=0.9\textwidth]{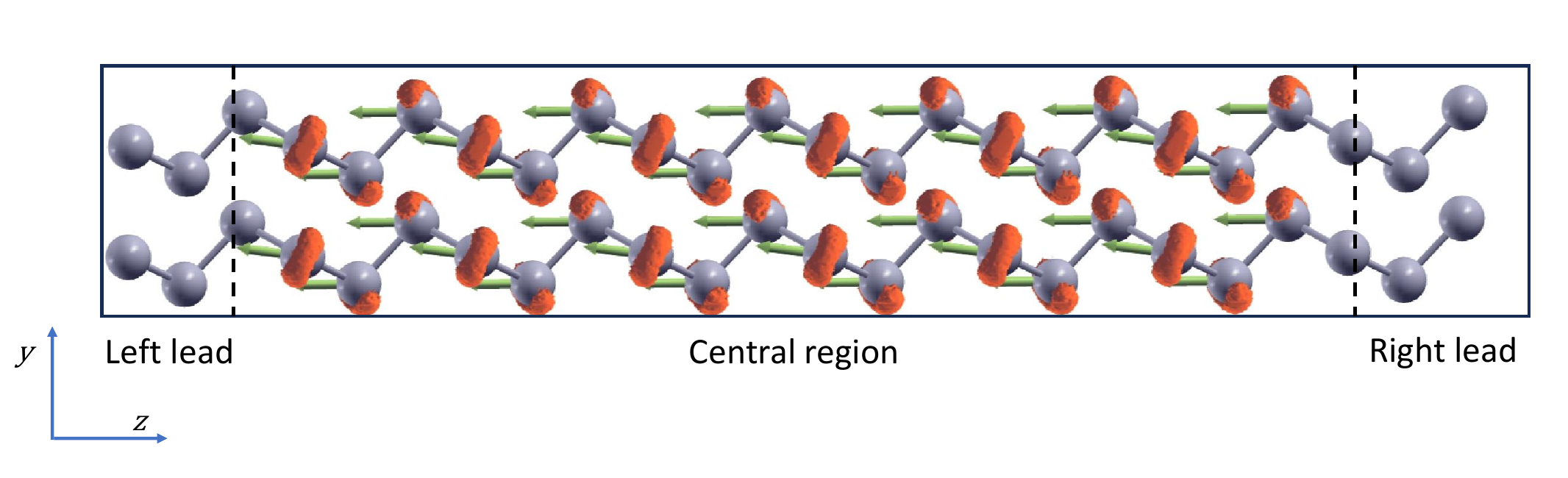}
     \caption{Typical systems studied in the DFT+NEGF calculations. It is subdivided into a central region (C), and left (L) and right (R) leads. The grey spheres are the Te atoms. The arrows represent the current induced magnetic moments. The orange ``bubbles'' mark the spin-z density isosurface.}\label{fig.CISP}
\end{figure*}

\section{System set-up, method and computational details}\label{sec.computational}

\subsection{DFT calculations}\label{sec_DFT_details}
The electronic structure of Te is obtained by using Kohn Sham DFT as implemented in a locally modified version of the SIESTA computational package \cite{soler2002siesta}. We perform the calculations for the hexagonal primitive unit cell of Te using the experimental lattice parameters listed in the previous section.
We assume the Perdew-Burke-Ernzerhof (PBE)\cite{PBE1,PBE2} generalized gradient approximation (GGA) for the exchange correlation functional. We include the SOC via the on-site approximation of Ref. [\onlinecite{fe.ol.06}].  
We treat core electrons with norm-conserving Troullier-Martins pseudo-potentials\cite{tr.ma.91_1,tr.ma.91_2} and expand the valence states through a numerical atomic orbital basis set including multiple-$\zeta$ and polarization functions. The cutoff radii of the basis orbitals are taken from Ref. [\onlinecite{ri.ga.15}]. We use a $51\times 51\times51\times$ Monkhorst-Pack $\mathbf{k}$ point grid to sample the Brillouin zone.

\subsection{DFT+NEGF transport calculations}\label{sec.DFT_NEGF_details}

We study transport and CISP along the Te chains by using DFT+NEGF calculations as implemented in the SMEAGOL code \cite{rocha2006spin,rungger2020non}, which is interfaced with the same SIESTA package used for the band structure calculations. The capability of SMEAGOL to describe materials with complex spin-textures was already demonstrated in Refs. [\onlinecite{na.ru.12,na.ru.dr.12,ja.na.15}]. The key features and equations used to study spin-charge conversion phenomena are presented in Ref. [\onlinecite{Bajaj}]. As such, in the following, we only give some computational details specific to the system and problems studied in this work, while we refer to that paper for a through discussion of the theoretical and technical aspects of the calculations.\\

We construct an infinite system along the $c$ axis, set parallel to the $z$ Cartesian direction, and we subdivide it in three parts: a central region, and left (L) and right (R) semi-infinite leads, from which electrons can flow in and out. Periodic boundary conditions are applied in the transverse $x$ and $y$ Cartesian directions. We assume that there is no disorder of any kind so that transport is ballistic.\\

Mathematically, the leads are represented through self-energies\cite{rungger2008algorithm} added to the DFT Kohn-Sham Hamiltonian of the central region. The Green's functions are then defined and evaluated following a standard procedure (see, for example, Refs. [\onlinecite{Datta, rocha2006spin,rungger2020non}]). We consider a rectangular supercell for the central region [see Fig. \ref{fig.CISP}-a and also Fig. S1-b in the SM]. Although we could equivalently consider a hexagonal supercell obtained by repeating the Te primitive unit cell, in practice, the use of a rectangular supercell allows for an easier calculation of quantities, such as the spin currents (see Refs. [\onlinecite{dr.to.22}] and [\onlinecite{Bajaj}]). Anyhow, the results in the following sections are presented in terms of real space quantities or converted in values per primitive unit cell, and therefore, the specific shape of the supercell used in the actual calculations is finally irrelevant.   \\ 

Each lead is effectively an electronic bath, characterized by the chemical potential, $\mu_{L}$ ($\mu_R$), and the temperature, $T_{L}$ ($T_R$). The central region is in thermodynamic equilibrium within the grand canonical ensemble when $\mu_R=\mu_L\equiv E_F$ and $T_L=T_R\equiv T$. In this case, the electrons are distributed according to the Fermi-Dirac distribution with Fermi energy, $E_F$, and temperature, $T$, and, as such, there is no flow of charge current between the leads. The equilibrium charge density is calculated self-consistently. In our case, since the central region is treated within the grand canonical ensemble, we simulate the p-doping in Te, by shifting $E_F$ to give any desired hole concentrations, and there is no need to add point defects. The Hartree potential, given by the Poisson equation, is matched between the various parts of the system (central regions and leads) to return the correct electrostatic profile and avoid divergences.\\

%The whole system is in thermodynamic equilibrium for $\mu_R=\mu_L\equiv E_F$ and  $T_L=T_R\equiv T$. The electrons in the central region are distributed according to the Fermi-Dirac distribution with Fermi energy $E_F$ and temperature $T$, and there is no flow of charge current. Then, 
The system can be driven out-of-equilibrium, thus inducing a charge current, by applying a bias voltage $V$ across the leads. This is practically done by displacing the leads' chemical potentials such that $\mu_L-\mu_R=eV\neq0$ ($e$ is the electron charge). Here, we use the rigid shift approximation \cite{Rudnev_Sci_Adv2017,xie2016spin}. The L (R) lead's chemical potentials and band structure are shifted by $+(-)eV/2$, and a linear ramp electrostatic potential $V(z)=-eVz/d+eV/2$ is added in the central region [$d$ is the length of the central region, and $z=0$ $(z=d)$ is the position of interface between the L(R) lead and the central region]. The finite bias calculations are then performed non-self-consistently, taking the zero-bias density matrix as input, and updating it only once. It is this update that leads to a slight modification of the density matrix compared to the equilibrium case and eventually gives rise to the spin-polarization in the driven system.\\

The charge and spin current values are obtained through the ``bond currents'' method\cite{Todorov_2002,Nikolic_2005_spin_acc, Theodonis_2006}, whose implementation is described in Refs. [\onlinecite{dr.to.22}] and [\onlinecite{Bajaj}]. We consider very small bias voltages and, by plotting the $I$-$V$ curve, we check that the system is in the linear-response limit. Our approach is then physically well justified, and, since we focus on the ballistic transport regime, the conductance $G=I/V$ is the Landauer conductance \cite{La.57}.\\

The CISP magnitude is estimated by calculating the real space spin density $\mathbf{s}(\mathbf{r})=[s^x(\mathbf{r}),s^y(\mathbf{r}),s^z(\mathbf{r})]$ whose definition is given in Eq. 6 of Ref. \onlinecite{Bajaj}. Alternatively, we compute the atomic magnetic moments $\mathbf{m}_\mathrm{i}=(m^x_\mathrm{i},m^y_\mathrm{i},m^z_\mathrm{i})$ from the Mulliken population analysis\cite{mu.55,Szabo} for each atom $i$. As discussed in Sec. \ref{sec.spin_density}, we find that the two approaches give comparable results in case of Te.\\ %In addition to the spin density and the magnetic moments, we can also compute the spin current, that represents the flow of spin. The definition of spin current within the used current approach is given in Ref. [\onlinecite{dr.to.22}].  \\ %of the SR can be identified with the Drude weight\cite{bo.go.04}, accounting for ballistic transport, in the linear-response Kubo formalism\cite{ga.ro.01}. \\ 

%The DFT+NEGF calculations are performed using the SMEAGOL code \cite{rocha2006spin,rungger2020non}, which is interfaced with the locally modified version of the SIESTA package that is used to obtain the electronic structure. The capability of SMEAGOL to describe materials with complex spin-texture was already demonstrated in Refs. [\onlinecite{na.ru.12,na.ru.dr.12,ja.na.15}]. \\

We employ the same exchange-correlation functional, basis set, and pseudopotential used for the DFT band structure calculations (see  Sec. \ref{sec_DFT_details}). 
We consider $31\times 31$ $\mathbf{k}$ points in the transverse Brillouin zone. The density matrix is calculated by integrating the lesser Green's function on a complex contour\cite{rocha2006spin, book2} using $16$ energy points along the complex semicircle, $16$ points along the line parallel to the real axis and $36$ poles. In the calculations at finite-bias, we specified $36$ additional points along the real energy axis for the integration of the non-equilibrium contribution to the density matrix. We thoroughly check that
the spin density is converged with respect to these calculation parameters. In particular, the convergence of the results with respect to the number of $\mathbf{k}$ points is discussed in Sec. S2 of the SM. Finally, we note that, since the calculated numerical values are small, our results are first obtained for a quite large current density ($\sim10^6$ A/cm$^{2}$) to achieve a reliable precision, and, afterwards, they are linearly extrapolated for the much lower current densities typically considered in experiments ($\sim10^2$ or $10^3$ A/cm$^{2}$). \\

\begin{figure*}[!ht]
    \centering
    \includegraphics[width=0.8\textwidth]{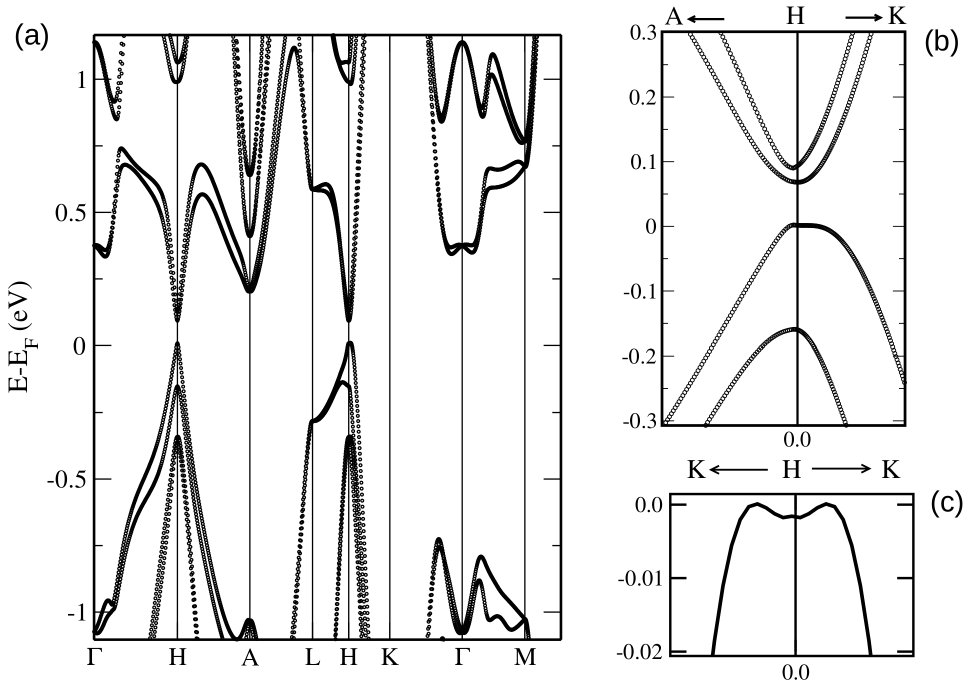}
    \caption{(a) Band structure of Te calculated by using Kohn-Sham DFT.  (b) Zoom around the $H$ high symmetry point in the Brillouin zone, showing the two topmost valence bands and the two lowest conduction bands. (c) Characteristic ``camel-back'' shape of the upper valence band with a local maximum along the $H$-$K$ line.}    
    \label{Fig.band_structure}
\end{figure*}

\section{Results}\label{sec.results}

\subsection{Band structure}\label{sec.bands}
We start by analyzing the electronic properties of Te obtained by using Kohn-Sham DFT, as detailed in Sec. \ref{sec_DFT_details}. Fig. \ref{Fig.band_structure} shows the band structure. The band gap is located at the $H$ high symmetry point in the Brillouin zone. In our calculations, its value, $\sim 80$ meV, is underestimated compared to the experimental value, 323 meV\cite{an.er.42}. This is a consequence of the well known ``band gap problem'' \cite{pe.le.83,sh.sc.83} of Kohn-Sham DFT and of the use of the PBE GGA functional. Recently, more accurate calculations were performed by means of the GW method \cite{sv.cr.10,hi.mo.15} and DFT with the hybrid functional HSE06 \cite{ts.pu.17}, obtaining quite accurate band gap predictions. However, the application of these approaches is not yet practically feasible in quantum transport, and, therefore, we need to rely on the GGA. In any case, the size of the band gap is practically irrelevant for the results that follow since we are only interested in p-doped systems.\\

When comparing our band structure with those obtained by using HSE06 or GW, we find that all the key features of the Te valence band are well reproduced. In particular, we recognize the characteristic ``camel-back'' shape of the upper valence band with a local maximum along the $H$-$K$ line (see Fig. \ref{Fig.band_structure}-c). Furthermore, we observe the two upper branches of the valence band, which are separated by 160 meV (Fig. \ref{Fig.band_structure}-b), in fair agreement with the HSE06 calculations of Ref. [\onlinecite{ts.pu.17}]. A few tenths of meV below the band edge around $H$, the spin texture (not shown) is radial in the $k_x$-$k_z$ plane
with spins pointing inward (outward) in right (left)-handed Te thus resembling a magnetic monopole in reciprocal space, like in the reports of Refs. [\onlinecite{ga.go.20, sa.hi.20}]. This is a signature of a prototypical Weyl SOC and leads to the unconventional nature of CISP in this material, as recently discussed in Refs. \onlinecite{Calavalle2022,Roy2022}.

\subsection{Non-equilibrium spin density} \label{sec.spin_density}

\begin{figure}
    \centering
    \includegraphics[width=0.4\textwidth]{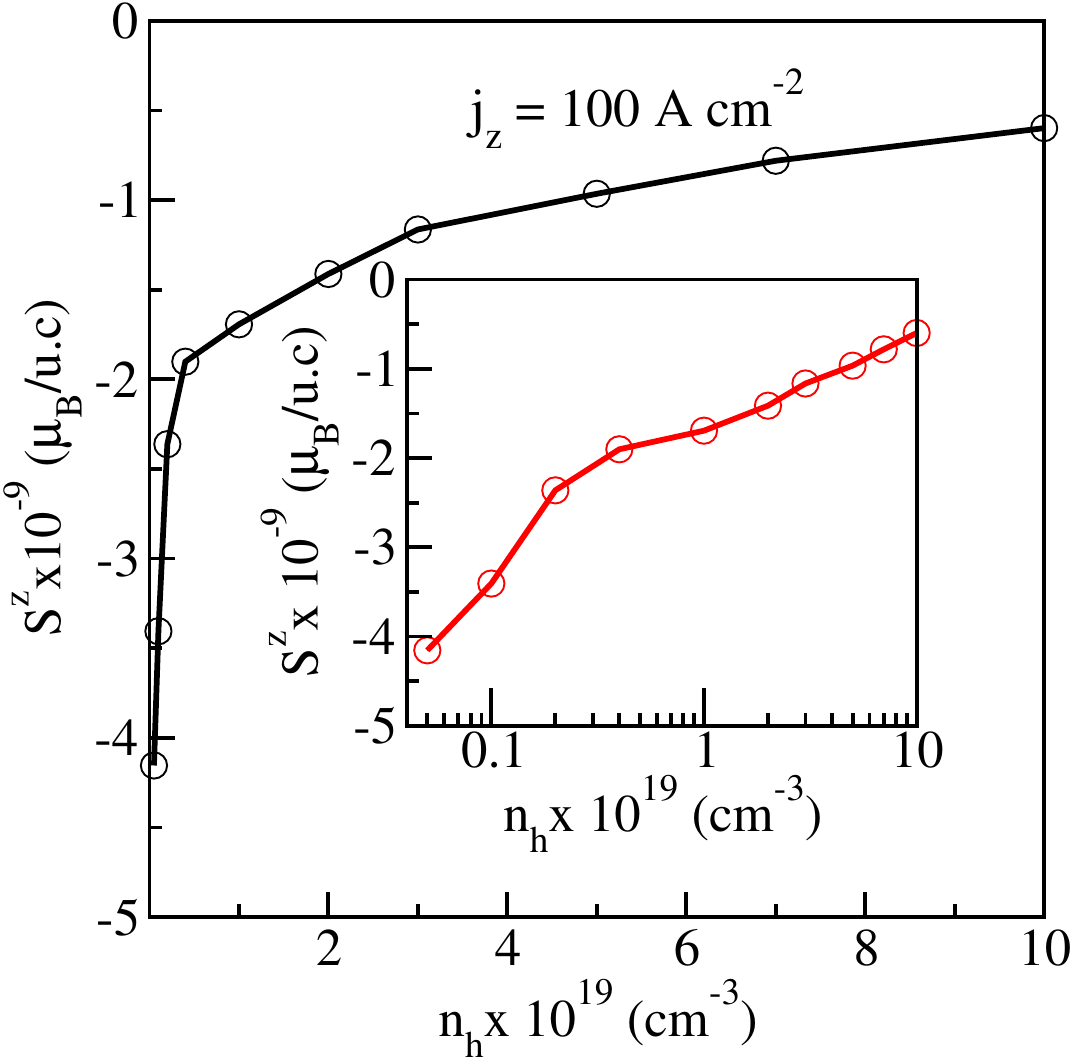}
    \caption{Spin density per unit cell, $S^z$, as a function of the hole concentration, $n_\mathrm{h}$. The results are for a charge current density, $j_z=100$ A/cm$^2$, and an electronic temperature, $T= 200$ K. The inset shows the same data, but with $n_\mathrm{h}$ plotted in logarithmic scale for a better display of the lowest doping region.}\label{Fig.spin_vs_density}
\end{figure}

\begin{figure}
    \centering
    \includegraphics[width=0.4\textwidth]{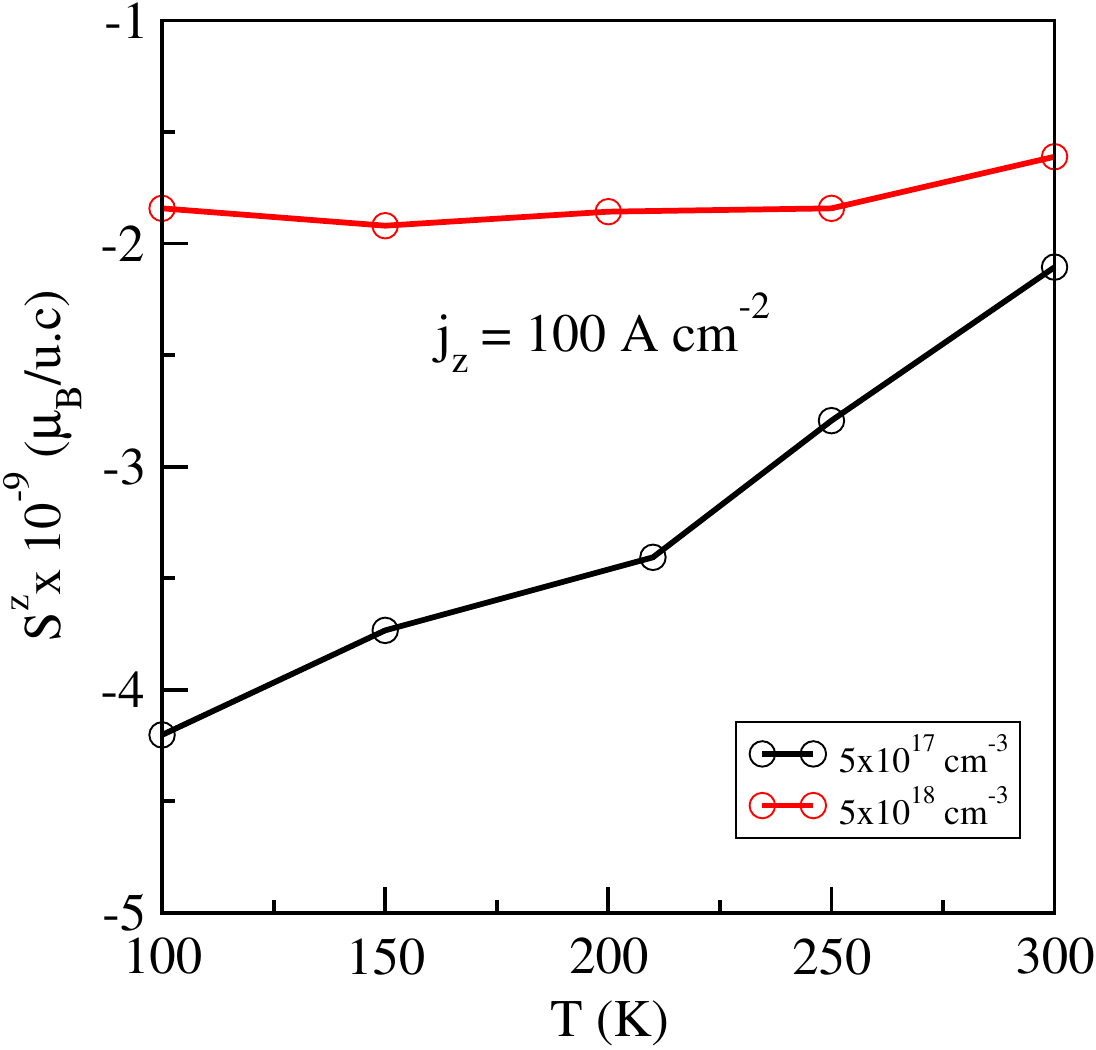}
    \caption{Spin density per unit cell, $S^z$, as a function of the electronic temperature, $T$. The results are for a charge current density, $j_z=100$ A/cm$^2$, and two different hole concentrations, $n_\mathrm{h}=5\times10^{17}$ cm$^{-3}$ (black dots) and $5\times10^{18}$ cm$^{-3}$ (red dots).}\label{Fig.spin_vs_temperature}
\end{figure}

%\begin{figure}
%    \centering
%    \includegraphics[width=0.4\textwidth]{V-T-points-removed.pdf}
%    \caption{\textcolor{red}{Confusing picture. Why do you plot such a large range. You need to zoom around the linear regime}\textcolor{cyan}{I noticed that the clustering of points was at very low values of the $\Delta T$ and it was not smooth graph when I zoomed in. That's why now I am keeping only one of those points and the range of $\Delta T$ used here is 20 to 150 K with the base Temp of 200K. Also, we need to mention that the black curve is for basis of 5mV with doping range between $5\times10^{17}$ to $5\times10^{19}$}. This is almost in the linear response regime when plotted in whole range, its due to zoom-in it looks a bit non-linear.}
%\end{figure}

We now study CISP in the p-doped Te chains by performing DFT+NEGF calculations, as detailed in Sec. \ref{sec.DFT_NEGF_details}.
The results presented in the following are %summarized in Fig. \ref{fig.CISP}
for the right-handed enantiomer, while those for the left-handed enantiomer are not shown as they can be trivially inferred from symmetry considerations.\\

In equilibrium, and therefore, in absence of a flowing charge current, the atomic magnetic moments are zero for all atoms inside our system. In contrast, when we apply a tiny bias voltage across the leads, small non-zero magnetic moments, shown as arrows in Fig. \ref{fig.CISP}-a, appear in the central region, eventually revealing the emergence of CISP. They are oriented in the direction parallel (antiparallel) to the current flow for left- (right-) handed Te, while their size increases linearly as a function of the current, in agreement with the model in Sec. \ref{sec.TeandCISP}. \\

CISP can be analyzed in a more detailed way in terms of the spin density in real space. $\mathbf{s}(\mathbf{r})$ is found to be zero everywhere inside the central region in equilibrium, whereas a non-zero component $s^z(\mathbf{r})$ appears for the driven system, meaning that the valence electrons get polarized along the $z$ transport direction. To visualize the effect we plot an isosurface of $s^z(\mathbf{r})$ in Fig. \ref{fig.CISP} (orange ``bubbles''). Most of the spin-z density appears localized around the atoms, thus confirming that the previous description in terms of magnetic moments is valid. However, the isosurface is not spherical. It is rather asymmetrical, presenting a lobe only on one side of each atom, and is perpendicular to the charge transport and spin-polarization direction $z$. Moreover, we note that it warps around a chain following the three-fold screw symmetry, $\mathbb{C}_3$. This means that the spin-z density at each atom is related to the one at the next atom along the chain through a 120-degree rotation around the $c$-axis. This can be further recognized by observing the same isosurface as in Fig. \ref{fig.CISP} from different perspective (see Fig. S5 in the SM). Overall, the plot of the spin-z density isosurface portraits the relation between spin and chirality in a very neat way. Our calculations provide a ``real space'' picture of CISP, complementary to the common description which accounts for the effect in terms of the spin texture in reciprocal space \cite{Roy2022,Calavalle2022}.

For a quantitative analysis and for a comparison with the other studies from the literature\cite{ts.pu.17, Roy2022}, we now 
calculate the spin density per primitive unit cell (u.c.), defined as $S^z= \int_{V_\mathrm{uc}} d\mathbf{r}\, s^z(\mathbf{r})$, where $V_\mathrm{uc}$ is the u.c. volume.
Alternatively, owing to the localization of the spin-z density near the atoms, the same quantity can can also obtained by summing the magnetic moments of the atoms inside the u.c., $S^z=\sum_{i\in N_\mathrm{uc}}m^z_i$.
We verified that the results from these two approaches are identical within two significant digits. They would instead be very different in systems where the spin-z density was mostly distributed in the interstitial region between the atoms\cite{Bajaj, dr.to.23}.\\

Since  CISP is an electron effect due to the out-of-equilibrium modification of the electron distribution, we consider only electronic temperature effects in our calculations, while the thermal displacement of the ions is neglected. The temperature ranges from $T=100$ to $300$ K, while calculations for lower temperatures can not be converged.
We also examine the effect of the doping by moving the Fermi energy position, as mentioned in Sec. \ref{sec.DFT_NEGF_details}. We vary the hole density, $n_\mathrm{h}$, %, from $10^{17}$ to $10^{20}$ holes/cm$^3$.
from $10^{17}$ cm$^{-3}$ to $10^{20}$ cm$^{-3}$, which are respectively the doping reported in Ref. [\onlinecite{Calavalle2022,Roy2022}] for Te and the largest doping achievable in semiconductors like Si. \\

Fig. \ref{Fig.spin_vs_density} displays $S^z$ as a function of the hole concentration for a charge current density $j_z=100$ A/cm$^{2}$. We observe a definite dependence of the results on $n_\mathrm{h}$. The absolute value $\vert S^z\vert$ is the largest $\sim 4.2 \times 10^{-9} \mu_B/$u.c. at the lowest considered doping, $n_\mathrm{h}=5 \times 10^{17}$ cm$^{-3}$. $\vert S^z\vert$ is then sharply reduced with increasing doping until its value eventually starts saturating for concentrations larger than $\sim 2 \times 10^{19}$ cm$^{-3}$. We can then conclude that a low hole concentration tends to maximize CISP. As already pointed out in Ref. [\onlinecite{Roy2022}], the reduction of $\vert S^z\vert$ with doping is due to the holes that start filling not just the topmost valence band, but also the second valence band, which has an opposite spin-texture. In this regard, we note that the Fermi energy reaches the edge of this second valence band for a concentration of about $\sim 2.5 \times 10^{19}$ cm$^{-3}$, that is indeed near where we observe a change in the slope of the $S^z$ versus $n_\mathrm{h}$ curve. \\

The temperature dependence of $S^z$ is presented in Fig. \ref{Fig.spin_vs_temperature}, again for a charge current density $j_z=100$ A/cm$^{2}$. In the case of a doping concentration $n_\mathrm{h}=5\times10^{17}$ cm$^{-3}$ (black dots), we observe that the absolute value of $S^z$ decreases with increasing $T$. Nonetheless, its order of magnitude does not change, i.e., it remains $ 10^{-9}\mu_\mathrm{B}/$u.c. At higher doping, the temperature dependence is completely washed out. For instance, in the case of $n_\mathrm{h}=5\times10^{18}$ cm$^{-3}$ (red dots), $\vert S^z\vert $ is essentially constant and equal to $1.8 \times 10^{-9}\mu_\mathrm{B}/$u.c. \\

Notably, our results are consistent with those from previous DFT calculations, even though they rely on different transport formalisms and exchange-correlation functionals. In Ref. [\onlinecite{ts.pu.17}], $\vert S^z\vert$ is predicted to be equal to about $ \sim 2 \times 10^{-9} \mu_B$ for a hole concentration $n_\mathrm{h}=10^{18}$ cm$^{-3}$ and a charge current density $j_z=100 $ A/cm$^{2}$. This is in good agreement with our estimate, which can be extracted from Fig. \ref{Fig.spin_vs_density}. A similar comparison can also be carried out with the results in Ref. [\onlinecite{Roy2022}]. Specifically, there, one can find a plot of $S^z$ as a function $n_\mathrm{h}$. The results in that plot look very similar to ours in Fig. \ref{Fig.spin_vs_density}, also displaying a comparable drop in the $\vert S^z\vert$ value when the doping is such that the Fermi energy reaches the edge of the second valence band. Overall, such good agreement between several DFT results is remarkable and provides a consistent quantitative picture of CISP in p-doped Te.\\

On the experimental side, we can refer to the transport study by Calavalle {\it et al.}\cite{Calavalle2022} and the NMR measurements by Furukawa {\it et al.}\cite{Furukawa2017}. The first did not provide any direct measure of $S^z$, which was nonetheless estimated via model calculations. The reported value is about $3\times10^{-8} \mathrm{\mu_B}$/u.c. for a doping concentration in the range from $7$ to $8\times10^{17}$ cm$^{-3}$ and a charge current density equal to 1000 A/cm$^2$. This result compares quite well with ours as we get $\vert S^z \vert \sim 3.5\times 10^{-8} \mathrm{\mu_B}$ for a similar doping and the same charge current density. In contrast, the NMR study provides a value that is an order of magnitude larger. Specifically, the reported spin density is equal to about $1.3\times 10^{-8}$~$\mathrm{\mu_B}$/u.c.
for a charge current density of $82$ A/cm$^2$.
However, we must remark that 
this estimate was indirect, rough, and tentative, as stated by the authors themselves. In particular, there are several problematic points. Firstly, the value of hyperfine coupling in the model used to fit the NMR data was not representative of the hyperfine coupling of the uppermost valence band, but rather it was the average value of those of the uppermost valence band and the conduction bands, which might be different. Secondly, the contribution from the orbital polarisation to the NMR shift was not  subtracted in the fit of the data thus possibly leading to an overestimation of the spin magnetic moment. Finally, we note that the NMR measurements were later repeated by the same group in a second paper\cite{PhysRevResearch.3.023111} obtaining a somewhat smaller current-induced shift. This would lead to a reduction in the magnetic moment value compared to the first paper, but unfortunately, the authors do not provide any updated estimates. Overall, further studies, beyond the scope of paper, may be required  to resolve the discrepancies between DFT and NMR estimates.

\subsection{CISP figure of merit}\label{sec.figure_merit}

CISP is often characterized by using a single parameter or ``figure of merit'' expressed by the ratio of the spin density over the charge current density \cite{to.kr.15,dr.to.23}. In the case of Te, this parameter coincides with the $\gamma_z^z$ component of the pseudotensor introduced in the phenomenological model of Sec. \ref{sec.TeandCISP}. Here, to calculate it in the same units as the ones considered in previous works, we first compute the so-called ``spin density profile'' \cite{dr.to.23}, that is the spin-$z$ density averaged over the $xy$ plane perpendicular to the current, $\langle s^{z}\rangle (z)=(l_xl_y)^{-1}\int_0^{l_x} \int_0^{l_y} dxdy\, s^{z}(x,y,z)$ with $l_x$ and $l_y$ the lengths of the supercell sides along the $x$ and $y$ directions (see Sec. S3 in the SM). Then, we define the spin density per atomic layer by integrating  $\langle s^{z}\rangle (z)$ over $z$ and dividing the result by the number of atomic layers, $N_L$, in the central region, i.e., $\langle S^z \rangle=\int dz \langle s^{z}\rangle (z)/N_L$. We then obtain that $\gamma\equiv\gamma_z^z= \langle S^z \rangle/j_z$ is $6 \times 10^3~ \mathrm{\mu_B}/$A for the doping concentration $n_\mathrm{h}=5\times10^{17}$ cm$^{-3}$ that maximizes CISP.\\

The results for Te can now be compared to the values for other materials studied in the literature. In particular, we consider the surfaces of heavy metals, often regarded as very efficient systems for spin-charge conversion\cite{Sanchez2013}. The $\gamma$ parameter of these surfaces was computed in Ref. [\onlinecite{dr.to.23}], where it was reported to be much larger for the early than it was for the late $5d$
materials. In the case of Te, our calculated $\gamma$ is comparable to the one of the W surface ($\sim 6 \times 10^3 \mathrm{\mu_B}/$A), which was predicted to be the best system overall. As such, Te turns out to be a similarly efficient CISP material. The advantage of Te compared to the $5d$ metals is however clear. CISP occurs in the bulk and is not just a surface effect.

\subsection{Spin current}\label{sec.spin_current}

\begin{figure}
    \centering
    \includegraphics[width=0.4\textwidth]{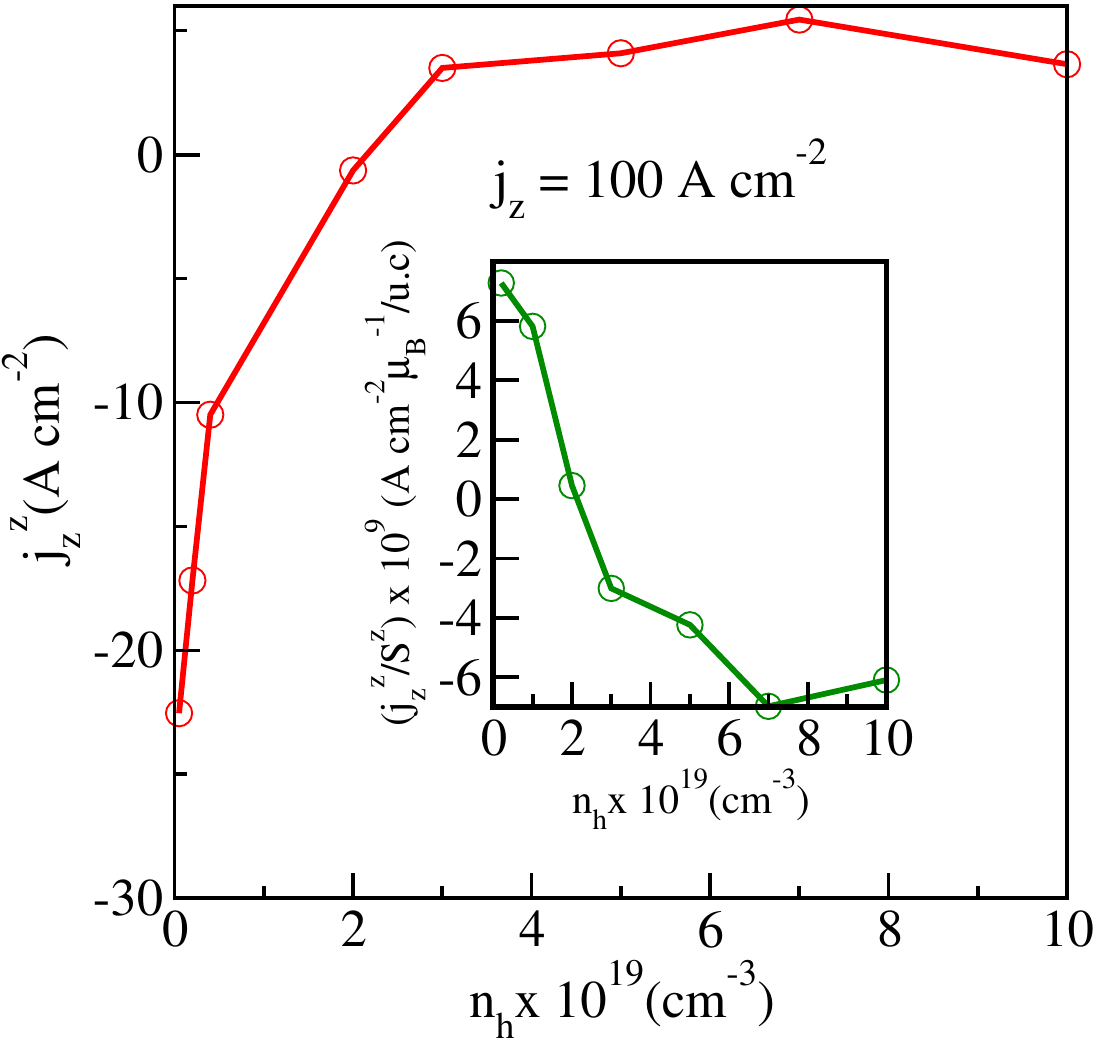}
    \caption{Spin current density, $j^z_z$, as a function of the hole concentration, $n_\mathrm{h}$. The results are for a charge current density $j_z=100$ A/cm$^2$ and an electronic temperature equal to $200$ K.  The inbox displays the ratio $\alpha=j^z_z/S^z$ as a function of the hole concentration.}\label{Fig.spin_current_vs_doping}
\end{figure}

The DFT+NEGF calculations show that the emergence of non-equilibrium magnetic moments in p-doped Te is also accompanied by a spin current density $j^z_z$ carrying spins polarized parallel to the transport direction, $z$. %In practice, here, the ratio of the spin current to the charge current densities, $P^z=j^z_z/j_z$, can be interpreted as the spin-polarization of the charge current. 
$j^z_z$ is defined, for example, in Ref. [\onlinecite{dr.to.22}] and is easily calculated by using the bond currents approach in the same way as done for the charge current. Nevertheless, we note that, in practice, some care is needed in the calculations. Because of the gyrotropic symmetry, Te has a large equilibrium (i.e., diamagnetic-like) spin current \cite{dr.to.22}, which can not be detected in experiments, but is returned by the calculations. This equilibrium spin current needs to be subtracted from the total spin current to obtain only the non-equilibrium component, which accounts for the transport of spins from one lead to the other and is therefore the spin current relevant for spintronics, as stressed by Rashba in Ref. [\onlinecite{ra.03}]. In the following, we will only provide values of the spin current density obtained after carefully subtracting the equilibrium part. \\

The computed spin current density $j^z_z$ is plotted in Fig. \ref{Fig.spin_current_vs_doping} as a function of hole concentration for the right-handed enantiomer. At the lowest considered doping, that is $n_\mathrm{h}\sim 10^{17}$,
$j^z_z$ is negative and its absolute value is maximum. It then sharply drops with increasing $n_\mathrm{h}$. Eventually at $n_\mathrm{h}\sim 2 \times10^{19}$ cm$^{-3}$, we observe a crossover with $j^z_z$ changing sign, i.e., switching from negative to positive, and saturating to a more or less constant value. Such crossover is due to the fact that, at $n_\mathrm{h}\sim 2 \times10^{19}$ cm$^{-3}$, the Fermi energy moves from the topmost valence
band to the second valence band, which has an opposite spin-texture, as already pointed out when discussing the $S^z$ versus $n_\mathrm{h}$ curve in Fig. \ref{Fig.spin_vs_density}. The results for the left-handed enantiomer would just have the opposite sign compared to those shown here, i.e., $j^z_z$ would be positive for hole concentrations below $n_\mathrm{h}\sim 2 \times10^{19}$ cm$^{-3}$ and would become negative for a larger doping. \\

To analyse the relation between the spin density and the spin current density, we plot the ratio $\alpha=j^z_z/S^z$ as a function of doping concentration and for a fixed charge current density (see the inset in Fig. \ref{Fig.spin_current_vs_doping}). When $n_\mathrm{h}$ is in the range from $5\times 10^{17}$ to $2\times10^{19}$ cm$^{-3}$, (i.e., before the cross-over) $\alpha$ has a more or less linear behavior. The emergence of the non-equilibrium spin density and of the spin current are two correlated effects that emerge together as a result of the Te spin texture. They can be seen as two complementary manifestations of CISP. \\ 

Finally, we can also compute the ratio of the spin current density to the charge current density, $P^z=j^z_z/j_z$. 
In practice for this work, $P^z$ can be interpreted as the spin-polarization of the charge current.
Remarkably, for $n_\mathrm{h}\sim 10^{17}$, that is when the CISP magnitude is maximum, we estimate that $P^z$ is about 0.24. This is a significant spin polarization for a material which is non-magnetic in equilibrium. Our calculations therefore fully support the expectation that Te can be used as an efficient material for all electrical spin generation and transport.\\

\subsection{Magnetoresistance}\label{sec.MR}

\begin{figure}
    \centering
    \includegraphics[width=0.38\textwidth]{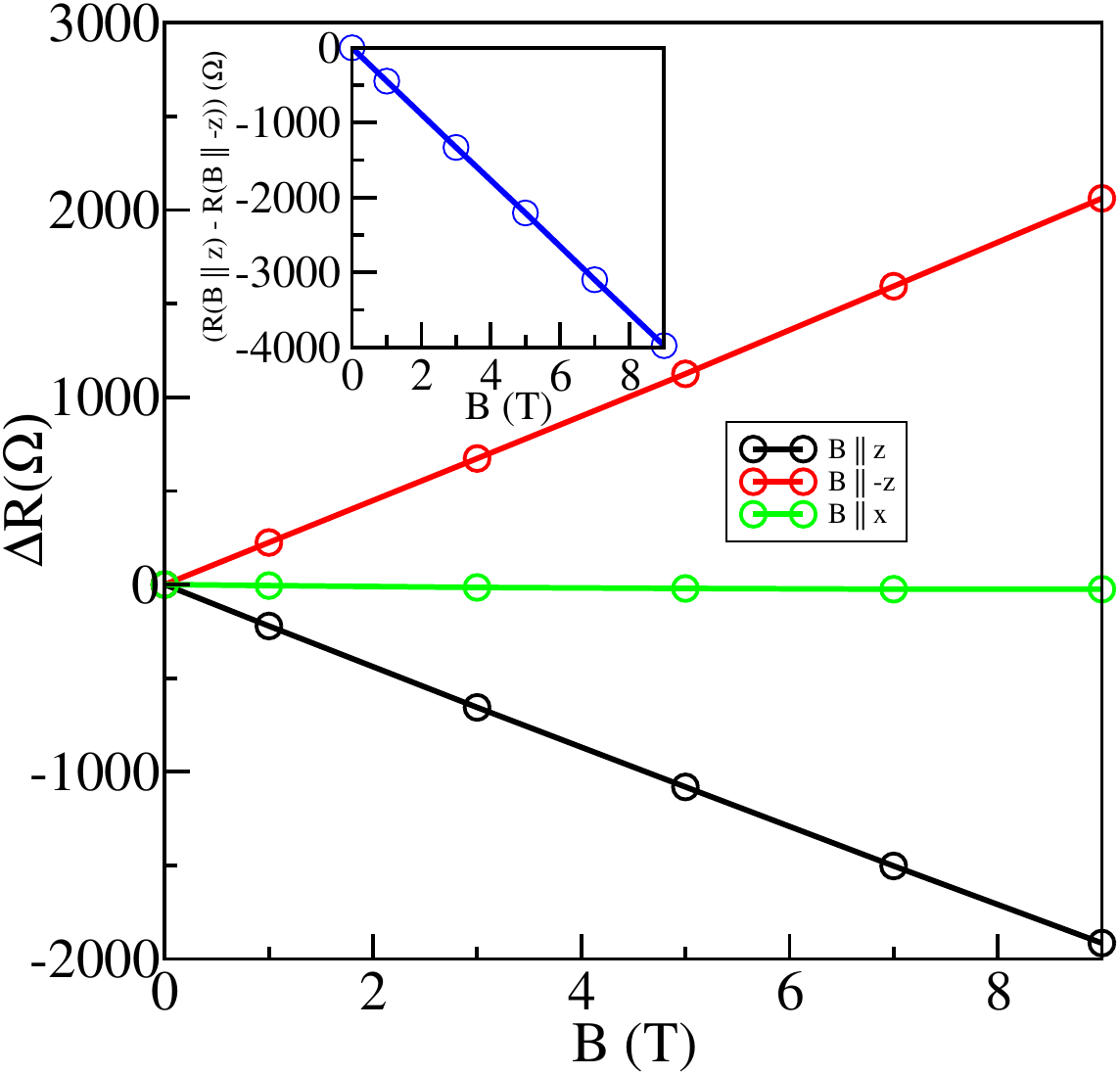}
    \caption{Resistance variation $\Delta R=R(\mathbf{B})-R(0)$ as a function of the magnetic field parallel, $\mathbf{B} \parallel \mathbf{z}$, antiparallel, $\mathbf{B} \parallel -\mathbf{z}$, and perpendicular, $\mathbf{B}\perp \mathbf{z}$ to the charge transport direction and for a positive bias. The results are for $n_\mathrm{h}=5\times 10^{17}$ cm$^{-3}$.}\label{Fig.DeltaR}
\end{figure}

\begin{figure}
    \centering
    \includegraphics[width=0.4\textwidth]{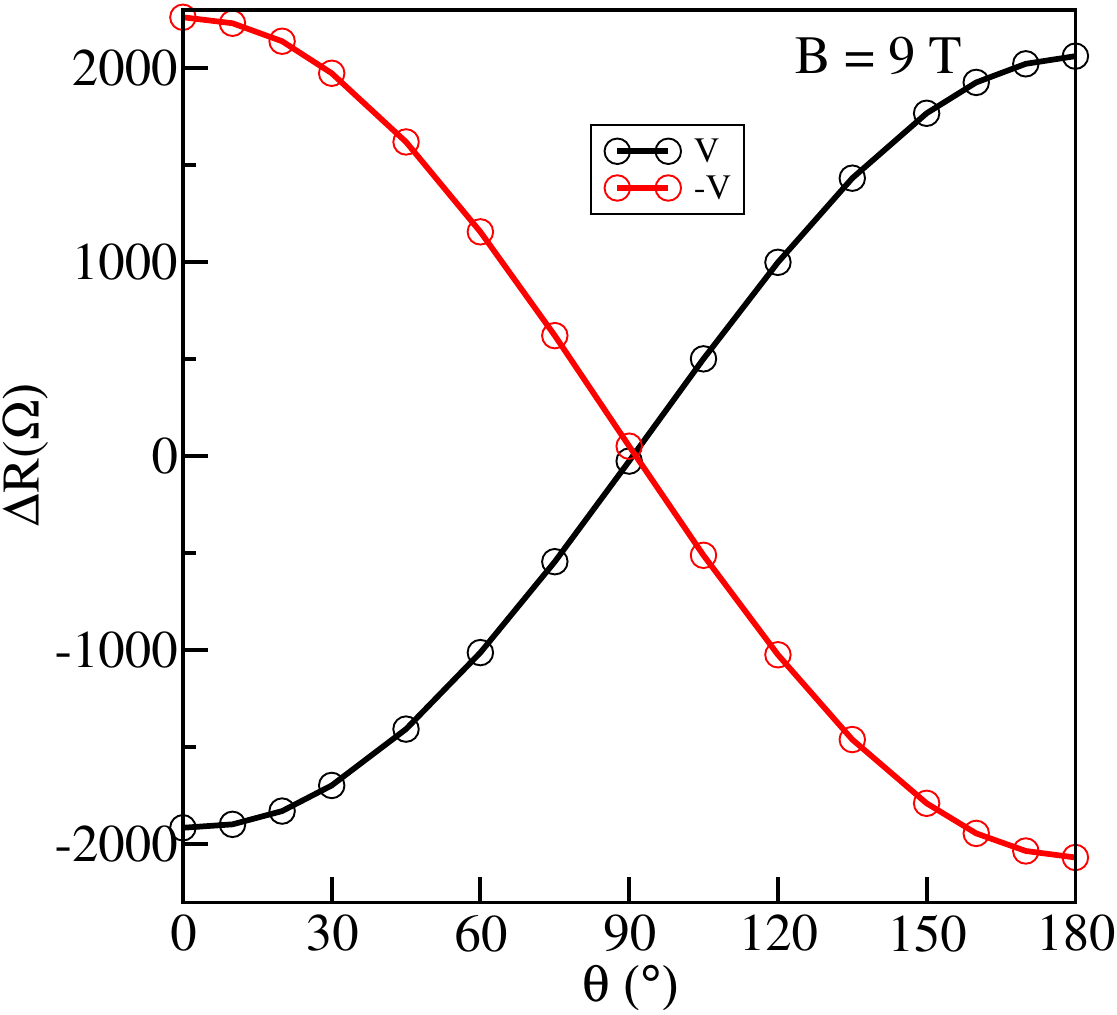}
    \caption{Resistance variation $\Delta R$ as a function of the magnetic field angle $\theta$ in the $xz$ plane. The black and red curves are obtained for positive and negative bias voltages, respectively.
    The results are for $n_\mathrm{h}=5\times 10^{17}$ cm$^{-3}$.}\label{Fig.R_vs_theta}
\end{figure}

\begin{figure}
    \centering
    \includegraphics[width=0.4\textwidth]{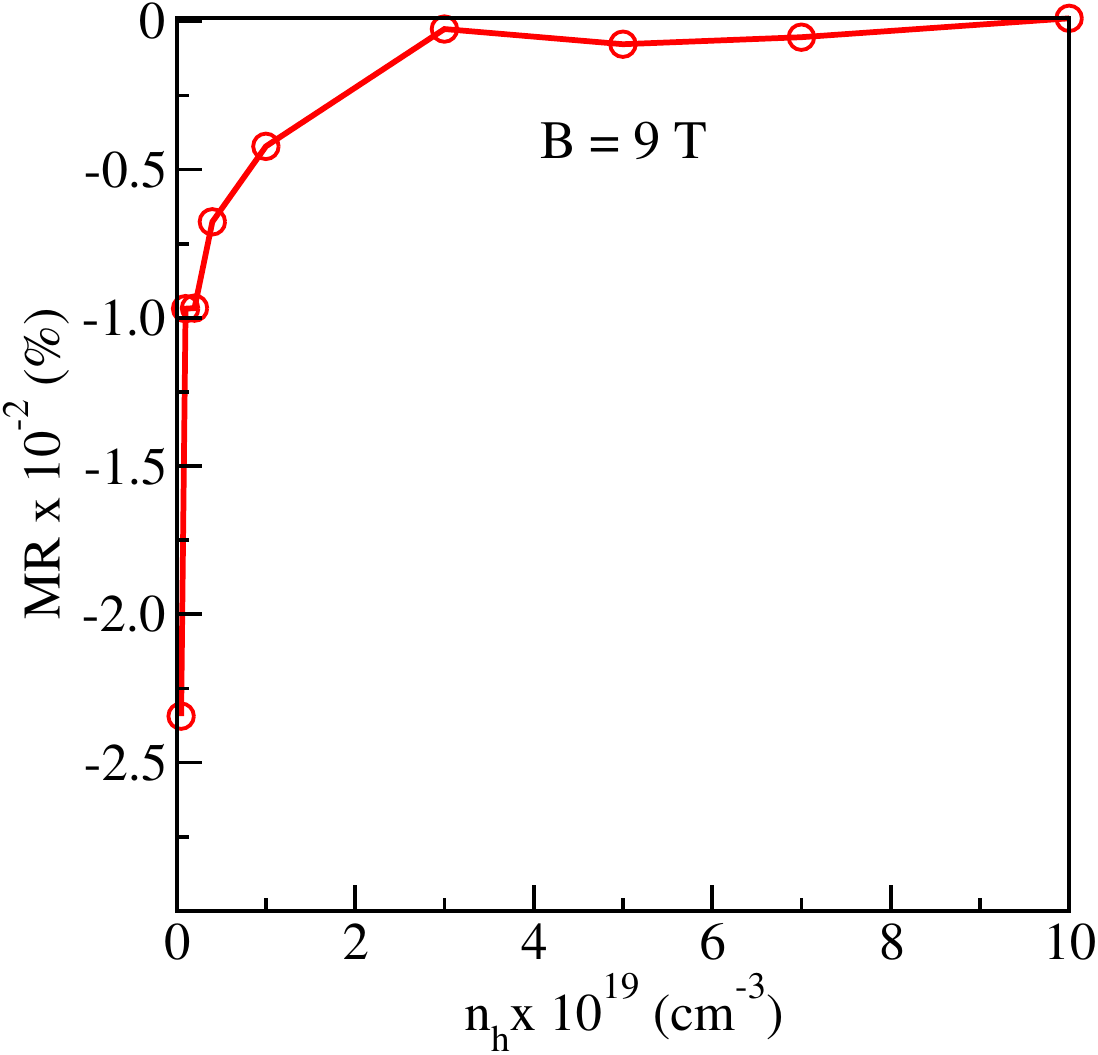}
    \caption{Effective current's spin polarization $P^z$ as a function of the magnetic field angle $\theta$ in the $xz$ plane. The results are for $5\times10^{17}$ cm$^{-3}$ and for the magnetic field modulus $B=9$ T. } \label{Fig.Pz_versus_theta}
\end{figure}

\begin{figure}
    \centering
    \includegraphics[width=0.4\textwidth]{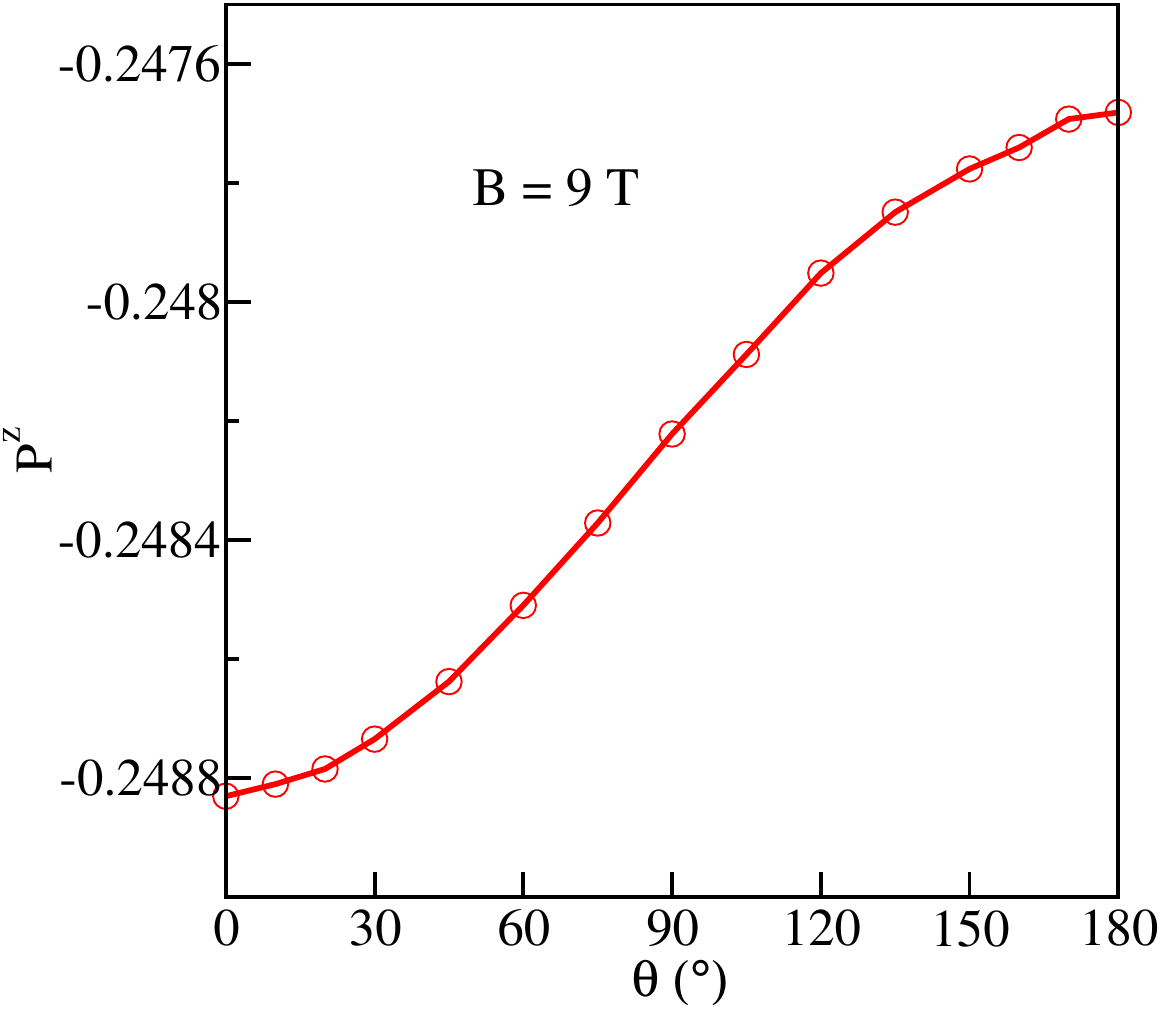}
    \caption{MR ratio for $B=9$ T as a function of the hole concentration $n_\mathrm{h}$.}\label{Fig.MR_ratio}
\end{figure}

%\begin{figure}
%    \centering
%    \includegraphics[width=0.4\textwidth]{pol-5-times-scaled.pdf}
%    \caption{\textcolor{red}{I would remove this picture.} \textcolor{cyan}{doping $5\times10^{17}$, T=200K, B = 9T, V=5mV but the graphs are shifted further by $2\times10^{-5}$ for better visualization. Also subtracted the small spin-pol at 90-degrees to make graphs more symmetric! Also, the graphs should be touching at 90-degrees if were not displaced and then they get more and more separated moving away from the center.}}
%\end{figure}

We now turn to the question of whether the unilateral magnetoresistance, or magnetochiral anisotropy, can appear in the ballistic transport regime as a result of CISP.
%Since CISP gives rise to an effectively spin-polarized current, we expect that it can be detected in magnetotransport measurements. 
To predict the effect at the quantitative level by using DFT+NEGF, we apply a Zeeman magnetic field $\mathbf{B}$ in the central region of our system, and we compute the resistance $R(\mathbf{B})$ when $\mathbf{B}$ is parallel, $\mathbf{B} \parallel \mathbf{z}$, antiparallel, $\mathbf{B} \parallel -\mathbf{z}$, and perpendicular, $\mathbf{B}\perp \mathbf{z}$ to the charge transport direction.
%To predict the effect at the quantitative level by using DFT+NEGF, we apply a Zeeman magnetic field $\mathbf{B}=(B\sin{\theta}, 0, B\cos{\theta})$ in the central region of our system, and we compute the resistance $R$ rotating $\mathbf{B}$ in the $xz$ plane to be parallel $\mathbf{B} \parallel \mathbf{z}$ ($\theta=0^\circ$), antiparallel $\mathbf{B} \parallel -\mathbf{z}$ ($\theta=180^\circ$) or perpendicular $\mathbf{B}\perp \mathbf{z}$ ($\theta=90^\circ$) to the charge transport direction.
%$\mathbf{B}=(B\sin{\theta}, 0, B\cos{\theta})$, which can be continuously rotated in the $xz$ plane to be parallel $\mathbf{B} \parallel \mathbf{z}$ ($\theta=0^\circ$), antiparallel $\mathbf{B} \parallel -\mathbf{z}$ ($\theta=180^\circ$) or perpendicular $\mathbf{B}\perp \mathbf{z}$ ($\theta=90^\circ$) to the transport direction, and we compute the resistance $R$. 
The resulting resistance variation with respect to the zero field resistance value, $\Delta R=R(\mathbf{B})-R(0)$, is shown in Fig. \ref{Fig.DeltaR} for the right-handed enantiomer with a hole concentration $n_\mathrm{h}=5\times 10^{17}$ cm$^{-3}$. In the perpendicular case (green dots), we observe that the resistance remains constant with increasing the modulus of the magnetic field, $B$. In contrast, when the magnetic field is parallel (antiparallel) to the direction of the current, the resistance monotonically decreases (increases) as a function of $B$ thus resulting in a negative (positive) magnetoresistance (see the red and black dots). The difference $R(\mathbf{B} \parallel \mathbf{z})- R(\mathbf{B} \parallel -\mathbf{z})$, shown in the inset of Fig. \ref{Fig.DeltaR}, is negative and varies linearly with $B$. The resistance is larger when the magnetic field is antiparallel to the current's direction. This is the unilateral magnetoresistance. Thus our calculations predict that indeed the effect can appear even in the ballistic transport set-up.
The unilateral magnetoresistance would be the opposite, i.e., larger for the parallel case, if we were considering the left instead of the right-handed enantiomer. \\ 

To further analyze the magnetoresistance, we now fix the modulus of the magnetic field at $B=9$ T (a typical value used in experiments\cite{Calavalle2022}) and rotate its direction in the $xz$ plane, $\mathbf{B}=(B\sin{\theta}, 0, B\cos{\theta})$, with $\theta=0^\circ$, $180^\circ$ and $90^\circ$ that respectively correspond to the parallel, antiparallel and perpendicular orientations, considered above. The change in the resistance as a function of $\theta$ is shown in Fig. \ref{Fig.R_vs_theta}(the same plot for the magnetic field rotated in the $yz$ plane is in Sec. S4 of the SM). The curve $\Delta R(\theta)$ has a $\sim\cos \theta$-like behaviour, characteristic of the electrical magnetochiral effect\cite{PhysRevLett.87.236602,at.tr.21}. The resistance depends linearly on the relative orientation of the magnetic field and of the charge current density, $\Delta R\sim \mathbf{B}\cdot\mathbf{j}$, with $\mathbf{j}=(0,0,j_z)$ in our set-up. 
The effect must be distinguished from the anisotropic magnetoresistance (AMR) in ferromagnetic materials, which displays a $\sim\cos^2 \theta$-like dependence and, therefore, an equal resistance for the magnetic field parallel and anti-parallel to the charge current direction. We note that, in our calculation we can only address the ``longitudinal'' magnetochiral effect where the current is along the Te wires. Rikken {\it et al.} in Ref. [\onlinecite{ri.av.19}] observed that the effect was maximized when the current was perpendicular to both the wires and the magnetic field. However, this result does not comply with the symmetry of the material\cite{liu2023electrical} and can not be related to its intrinsic properties, regardless of the implied microscopic mechanisms.\\

%The magnetoresistance ratio, $\mathrm{MR}=[R(\theta=0^\circ)-R(\theta=180^\circ)]/R(\theta=0^\circ)$ is plotted in Fig. \ref{Fig.MR_ratio} as a function of the hole concentration, $n_\mathrm{h}$, and for the magnetic field modulus, $B=9$ T. The trend closely resembles that of $S^z$ and $j_z^z$ in Figs. \ref{Fig.spin_vs_density} and \ref{Fig.spin_current_vs_doping}, further demonstrating the existing correlation between CISP and transport properties. The absolute value of the $\mathrm{MR}$ ratio is maximum, for the lowest considered doping, $n_\mathrm{h}=5\times10^{17}$ cm$^{-3}$, but it is unfortunately small ($0.025\%$) and probably not measurable in potential experiments. The MR then sharply drops for larger hole concentrations, and finally vanishes when the Fermi energy moves away from the top valence band and crosses the second valence band, i.e., for $n_\mathrm{h} > 2\times10^{19}$ cm$^{-3}$.\\ 

The magnetotransport characteristics of our system can be related to the microscopic quantities introduced in the previous sections.
For instance, in Fig. \ref{Fig.Pz_versus_theta}, we display the effective current spin polarization, $P^z=j^z_z/j_z$ as a function the magnetic field angle $\theta$ for a hole concentration, $n_\mathrm{h}=5\times 10^{17}$ cm$^{-3}$ and $B=9$ T. $P^z(\theta)$ follows a $\sim \cos{\theta}$-like curve similar to that of the resistance, $R(\theta)$. The absolute value of $P^z$ is enhanced (reduced) when the magnetic field is parallel (antiparallel) to the current, $\theta=0^\circ$ ($\theta=180^\circ$). This indicates that the magnetoresistance serves to sense the current spin polarization.\\

The emergence of the unilateral magnetoresistance in our calculations is a consequence of two combined factors, the Te chirality and the out-of-equilibrium driving bias, which are both essential. If the system was not chiral, i.e., if it presented a longitudinal plane of symmetry, the resistance would be the same with the magnetic field parallel or anti-parallel to the transport direction, as discussed, for instance, in Ref. [\onlinecite{de.ga.23}]. On the other hand, even for chiral systems, linear response calculations without computing the out-of-equilibrium density would give no magnetoresistance because of Onsanger's reciprocity, as demonstrated in Refs. [\onlinecite{zh.fe.05,ga.bl.23}]. Osanger's reciprocity no longer holds in our finite bias calculations, and, therefore, the magnetoresistance is allowed. %, and, as a consquence of chirality, the resistance changes for the magnetic field parallel and anti-parallel to the transport direction. 
The relation of the magnetoresistance with the system's chirality can be further verified by reversing the voltage $V\rightarrow -V$. When doing so we find that $R(\theta)\rightarrow R(\theta+180^\circ)$ (compare the black and red curves in Fig. \ref{Fig.R_vs_theta}). Similarly, if we selected left-handed instead of right-handed Te, we would find the reversal of the effect.\\

%Recently, the electrical magnetochiral effect was measured in Te nanowires by Calavalle {\it et al.} \cite{Calavalle2022}, as already mentioned in the introduction. Like in our calculations, they found that  the resistance significantly changes when the magnetic field is switched from parallel to antiparallel to the charge current direction. However, it is important to stress that the magnetoresistance studied here is distinct from that reported in the experiment. In our case, it emerges in the ballistic transport regime, whereas the experiment considers very resistive samples. A theory for the electrical magnetochiral effect based on the semiclassical Boltzmann formalism, valid for ohmic transport, was derived by Liu {\it et al.} in Ref. [\onlinecite{liu2023electrical}], which can be considered as a complementary work to ours. Interestingly, prior to the work by Calavalle {\it et al.}, the electrical magnetochiral effect was reported in Te by Rikken {\it et al.} in Ref. [\onlinecite{ri.av.19}]. However, these authors observed that the effect was maximized when the magnetic field was perpendicular rather than parallel to the current. This result does not comply with the symmetry of the material\cite{liu2023electrical} and can not be related to the intrinsic properties of Te, regardless of the implied microscopic mechanisms.\\

To estimate the unilateral magnetoresistance we can use two different quantities, namely the magnetoresitance ratio and the
magnetochiral anisoptropy parameter\cite{ri.av.19,liu2023electrical}. The magnetoresitance ratio is defined as $\mathrm{MR}=[R(\theta=0^\circ)-R(\theta=180^\circ)]/R(\theta=0^\circ)$. It is plotted in Fig. \ref{Fig.MR_ratio} as a function of the hole concentration, $n_\mathrm{h}$, and for the magnetic field modulus, $B=9$ T. %The trend closely resembles that of $S^z$ and $j_z^z$ in Figs. \ref{Fig.spin_vs_density} and \ref{Fig.spin_current_vs_doping}, further demonstrating the existing correlation between CISP and transport properties. 
The absolute value of the $\mathrm{MR}$ ratio is maximum, albeit small ($0.025\%$), for the lowest considered doping, $n_\mathrm{h}=5\times10^{17}$ cm$^{-3}$. The MR then sharply drops for larger hole concentrations, and finally vanishes when the Fermi energy moves away from the top valence band and crosses the second valence band, i.e., for $n_\mathrm{h} > 2\times10^{19}$ cm$^{-3}$. The trend closely resembles that of $S^z$ and $j_z^z$ in Figs. \ref{Fig.spin_vs_density} and \ref{Fig.spin_current_vs_doping}, further demonstrating the existing correlation between CISP and transport properties.\\ 

The magnetochiral anisoptropy parameter, $\alpha$, is defined through the ratio \cite{ri.av.19,liu2023electrical}
\begin{equation}
\frac{R(B,j_z)-R(B,-j_z)}{R(B,j_z)+R(B,-j_z)}=2\alpha Bj_z,
\end{equation}
where $R(B,j)$ and $R(B,-j)$ are the resistance for the same magnetic field, parallel to $\mathbf{z}$, and for opposite charge current densities of equal modulus $j$. In our case, a fit of the numerical data gives $\alpha$ of the order of $10^{-13}$ to $10^{-14}$ m$^2$/AT.\\ %Interestingly, beside this longitudinal the electrical magnetochiral effect was first reported in Te by Rikken {\it et al.} in Ref. [\onlinecite{ri.av.19}]. However, these authors observed that the effect was maximized when the magnetic field was perpendicular rather than parallel to the current. This result does not comply with the symmetry of the material\cite{liu2023electrical} and can not be related to its intrinsic properties, regardless of the implied microscopic mechanisms.\\ 

Our calculated parameter $\alpha$ is three order of magnitude smaller than that obtained in Ref. [\onlinecite{ri.av.19}]. Furthermore, the small magnetoresistance ratio reported in our ballistic transport calculations contrasts with the results reported by Calavalle {\it et al.} in the ohmic transport regime, where the unilateral magnetoresistance can reach almost $10\%$. These findings suggest that, although the electrical magnetochiral effect is intrinsic to the material, scattering with defects becomes essential to observe it. In support of this argument, we note that, in experiments, the effect is enhanced with increasing the resistivity of the samples. As such, further theoretical studies, including disorder and bridging from the ballistic to the ohmic limit will be essential to fully clarify this behavior, but unofrtunately, they remain practically challenging because of the large computational overhead.  \\ 

In summary, we have shown that CISP in ballistic Te wires leads to the formation of magnetic moments around the atoms as well as spin currents, which can be sensed by measuring a unilateral, albeit very small, magnetoresistance. Overall, our results provide a rather complete view of the phenomenon at an accurate quantitative level. Next, we will see show how the reported physics is evocative of the one emerging in quantum transport calculations for chiral molecular junctions\cite{na.mu.23, ga.bl.23}.

\section{Discussion: CISP and CISS within the DFT+NEGF framework}\label{sec.discussion}

Several recent works relied on DFT+NEGF calculations to study chiral molecular junctions in the attempt to shed some light on the CISS effect. The method implementations, the systems' set-up (comprising a central region attached to leads), and the numerical analysis were quite similar to ours here (see, for example, the computational method section in Ref. [\onlinecite{ga.bl.23}] and compare it with our paper, Ref. [\onlinecite{Bajaj}]). It is therefore interesting to look at the results to understand whether the predicted behaviours of chiral molecular junctions and chiral Te are related. %the similarities between the predicted behaviour of chiral molecular junctions and chiral Te. 
We identify a few key features common to both systems.\\

\begin{itemize}
\item[1 -] There are magnetic moments appearing around the atoms in finite bias calculations. They stem from the out-of-equilibrium component of the spin density. This effect is graphically represented in Fig. \ref{fig.CISP} here for Te and in Fig. 1 in Ref. [\onlinecite{ga.bl.23}] or in Fig. 4 in Ref. [\onlinecite{na.mu.23}] for several molecular junctions. In Te, the magnetic moments are collinear, parallel to the transport direction, whereas in molecular junctions they are not. Yet, this difference can be simply attributed to the different systems' symmetry.  \\

\item[2 -] A spin current emerges alongside with the magnetic moments. We showed it in the case of Te via a direct calculation in Sec. \ref{sec.spin_current}. In contrast, studies about molecular junctions did not explicitly evaluate such spin current. Yet, they present the spin-polarization of the energy-dependent transmission coefficient (see Fig. 2 in  Ref. [\onlinecite{na.mu.23}]). Within the Landauer-B\"uttiker formalism\cite{La.57,Bu.86,Bu.88}, implied in those calculations, the spin current is roughly given by the integral of such transmission coefficient over an energy range equal to the difference of the leads' chemical potentials (see, for example, Ref. [\onlinecite{PhysRevB.79.094414}]). If the spin-polarization of the transmission is non-zero, even the spin current will evidently be non-zero.\\ 

\item[3 -] The spin current leads to a unilateral magnetoresistance effect. On the one hand, to probe it, we applied an external magnetic field, $\mathbf{B}$, in the Te central region. On the other hand, calculations for molecular junctions relied on a ferromagnetic electrode, whose magnetization, $\mathbf{M}$, was rotated along different directions. Yet, the qualitative results are finally the same. In Te, the current spin-polarization in Fig. 9 changes when the direction of the magnetic field is fliped by $180^0$, i.e. we get $\Delta P(\mathbf{B})=P^z(\mathbf{B})-P^z(-\mathbf{B})\neq 0$. Similarly, in chiral molecular junctions, one finds that there is a non vanishing change in the current spin polarization, $\Delta P(\mathbf{M})\neq 0$, when reversing $\mathbf{M}$, as shown in Fig. 2 of Ref. [\onlinecite{ga.bl.23}] (purple dotted lines).
\end{itemize}

The comparison of the results between chiral molecular junctions and Te can also be extended from a qualitative to a quantitative level. In particular, we note that Te has a larger SOC than molecular junctions made of lighter atoms, mostly C. Consistently, the maximum spin polarization estimated in Ref. [\onlinecite{ga.bl.23}] for a molecule is $<1\%$, whereas, in Te, we obtain that $P^z$ can reach about $24\%$. This difference of three orders of magnitude appears to follow the ratio of the atomic SOC of Te and of C, i.e. $\sim (Z_\mathrm{Te}/Z_\mathrm{C})^4$.\\

Summarizing, from our analysis, we can not observe any qualitative differences in the results of DFT+NEGF calculations for chiral molecular junctions and Te. This leads us to an important conclusion. The emergence of non-equilibrium magnetic moments, spin currents and magnetoresistance, that, in previous works, were considered fingerprints of CISS, can alternatively and indifferently be attributed to CISP, as we have done in this paper. There is no way to distinguish between the two effects from the results of DFT+NEGF calculations. In other words, we can say that the two effects are the same {\it within the framework of DFT+NEGF}. However, we note that a distinction between the two effects may still appear when considering other situations, namely polaronic transport in long organic molecules \cite{PhysRevB.102.214303}, and the ohmic transport regime in solid state materials \cite{Calavalle2022, Belashchenko}. A careful analysis of this problem is however beyond the scope and limitations of this paper. \\

Finally, we note that the analogy between Te and molecular junctions may also be interesting in order to find out possible routes to enhance the unilateral magnetoresistance. In molecules, the inclusion of inelastic effects is believed to be critical for enhancing the CISS effect \cite{du.hu.20,fr.20,fr.22,hu.ka.21}. Similar considerations may also apply to Te. In this regard, DFT+NEGF calculations can, in principle, be extended to include electron-electron\cite{dr.ra.22} or electron-phonon\cite{fr.pa.07} scattering via self-energies. More work along this direction is currently on going, and the results will further boost our understanding of CISP and CISS across different systems. 

\section{Conclusion}\label{sec.conclusion}
In this paper we studied CISP in p-type Te via DFT+NEGF calculations, focusing on the ballistic transport regime. We found that the effect could be quantitatively described in terms of two complementary quantities, namely the non-equilibrium atomic magnetic moments and the spin current polarized parallel to the transport direction.\\

The estimated values of the magnetic moments and their dependence on the hole concentration are in good agreement with those from previous DFT studies\cite{ts.pu.17,Roy2022}. The obtained CISP figure of merit, that is the
ratio of the spin density over the charge current density,
is comparable to that of the well-studied $5d$ metal thin films, indicating that Te is a similarly efficient charge-to-spin conversion material. The advantage of Te compared to the $5d$ metals is however that CISP occurs in the bulk and is not just a surface effect.\\

The spin current is estimated to be about $24\%$ of the charge current. This a substantial fraction for a material which is non-magnetic in equilibrium. To our knowledge the spin current associated to CISP was not previously calculated for any system.\\

The spin current can in principle be detected via magnetotransport measurements. Specifically, we predict the existence of a ballistic unilateral magnetoresistance, or electric magnetochiral anisotropy, seen as a change in the resistance of a Te wire when an external magnetic field is switched 
from parallel to antiparallel to the direction of the charge current. The calculated $\mathrm{MR}$ is however small, about $0.025\%$.\\
%Specifically, we predict that the resistance along a Te wire changes when an external magnetic field is applied parallel or antiparallel to the charge current direction. The effect is a signature of the chirality of the material and is the ballistic transport analogous of the electrical magnetochiral anisotropy, reported in recent experiments \cite{Calavalle2022}. The calculated $\mathrm{MR}$ is however quite small, about $0.025\%$.\\
%Finally, the description of CISP in terms of spin current allows us to establish a connection with recent DFT+NEGF calculations for chiral molecular junctions. We conclude that CISP and CISS are the same effect within a pure ballistic picture of transport.   \\

Finally, we compared our work for Te with recent DFT+NEGF calculations for chiral molecules. We found that the results were comparable with differences that could be ascribed to specific material properties or symmetry. In other words, we were not able to identify any characteristic features, which could be used to distinguish CISS from CISP. Hence, we conclude, that, within the framework of DFT+NEGF, the two effects are the same.\\

%\section{equations}
%\begin{equation}
%\langle s^{z}\rangle (z)=\frac{1}{l_xl_y}\int_0^{l_x} \int_0^{l_y} dxdy\, s^{z}(x,y,z)
%\end{equation}
%(this should be the quantity outputted by the pot.sh script).
%where $l_x$ and $l_y$ are the lateral size of the central region supercell.\\
%The total spin density per supercell is
%\begin{equation}
% S^{z}_{S.C.}=(l_xl_y)\int_0^{L} dz\, \langle s^{z} \rangle (z)
%\end{equation}
%where $L$ is the length of the central region. This means that you take fig. 1 and you integrate $\langle s^{z} \rangle (z)$ over $z$ and then you multiply it by $l_xl_y$.\\
%The spin density per unit cell is then given as
%\begin{equation}
%S^z_{u.c.}=   S^{z}_{S.C.} V_{u.c.}/V_{S.C.} 
%\end{equation}
%where $ V_{u.c.}$ is the volume of the premitive unit cell, and $V_{S.C.}$ is our central region volume.
%Is this the quantity that you are calculating?\\
%Is this the quantity that is calculated in the literature?\\
%Alternatively we can sum the atomic magnetic moments $m^z_I$ from the mulliken population 
%\begin{equation}
%M^z=\sum_{I\in N} m_I^z
%\end{equation}
%where $N$ is the number of atoms in the central region.
%Then per unit cell
%\begin{equation}
%M^z_{u.c.}=M^z N_{uc}/N
%\end{equation}
%where $N_{uc}$ is the number of atoms in the Te primitive unit cell.

\begin{acknowledgments}
This work was sponsored by Science Foundation Ireland and the Royal Society through the University Research Fellowships URF-R1-191769, and by the European Commission through the H2020-EU.1.2.1 FET-Open project INTERFAST (project ID 965046). The computational resources were provided by Trinity College Dublin Research IT and the Irish Centre for High-End Computing (ICHEC). 
\end{acknowledgments}

\end{document}